\documentclass[prb,aps,superscriptaddress,twocolumn,showpacs]{revtex4}
\usepackage{graphicx,amsfonts,bm,amsmath}
\newcommand{\bk}{b_{\bm{k}}}

\newcommand{\bmkd}{b_{-\bm{k}}^{\dag}}
\newcommand{\bek}{\beta_{\bm{k}}}
\newcommand{\bemk}{\beta_{-\bm{k}}}
\newcommand{\bekd}{\beta_{\bm{k}}^{\dag}}
\newcommand{\sumk}{\sum_{\bm{k}}}
\newcommand{\wk}{\omega_{\bm{k}}}

\newcommand{\kk}{\bm{k}}
\newcommand{\rl}{\rangle\!\langle}

\DeclareMathOperator{\tr}{Tr}

\begin{document}

\author{A. Grodecka}
\email{anna.grodecka@pwr.wroc.pl}
\affiliation{Insitut f{\"u}r Theoretische Physik,
Nichtlineare Optik und Quantenelektronik,
Technische Universit{\"a}t Berlin, 10623 Berlin, Germany}
\affiliation{Institute of Physics, Wroc{\l}aw University of Technology,
50-370 Wroc{\l}aw, Poland}
\author{C. Weber}
\affiliation{Insitut f{\"u}r Theoretische Physik,
Nichtlineare Optik und Quantenelektronik,
Technische Universit{\"a}t Berlin, 10623 Berlin, Germany}
\author{P. Machnikowski}
\affiliation{Institute of Physics, Wroc{\l}aw University of Technology,
50-370 Wroc{\l}aw, Poland}
\author{A. Knorr}
\affiliation{Insitut f{\"u}r Theoretische Physik,
Nichtlineare Optik und Quantenelektronik,
Technische Universit{\"a}t Berlin, 10623 Berlin, Germany}

\title{Interplay and optimization of decoherence mechanisms\\
in the optical control of spin quantum bits\\
implemented on a semiconductor quantum dot}

\begin{abstract}
We study the influence of the environment on an optically induced
rotation of a single electron spin in a charged semiconductor quantum dot.
We analyze the decoherence mechanisms resulting from the dynamical lattice 
response to the charge evolution induced in a trion-based optical spin 
control scheme. Moreover, we study the effect of the finite trion
lifetime and of the imperfections of the unitary evolution
such as off-resonant excitations and the nonadiabaticity of the driving.
We calculate the total error of the operation on a spin-based qubit
in an InAs/GaAs quantum dot system and discuss possible optimization
against the different contributions.
We indicate the parameters which allow for coherent control of the spin
with a single qubit gate error as low as $10^{-4}$.
\end{abstract}

\pacs{63.20.Kr, 03.65.Yz, 03.67.Lx, 42.50.Hz, 71.35.Pq}

\maketitle

\section{Introduction}

The spin degree of freedom of an excess electron
in a charged semiconductor quantum dot (QD) has been proposed
as an attractive candidate for the use as a qubit
\cite{calarco03,imamoglu99,feng03}.
The advantage of storing the logical values in spin states is
their long coherence time \cite{hanson03}.
In addition, it is possible to optically induce charge dynamics dependent
on the spin state of an electron via the Pauli exclusion principle
and optical selection rules \cite{pazy03,emary07,economou06}.
Coupling between charge states via static \cite{pazy03,calarco03} 
and interband \cite{nazir04,lovett05} electric dipole moments allows
one to perform quantum conditional operations on two spins. In this
way, one can implement single-  and two-qubit gates in QD
systems. 
The utilization of the optical control methods leads to shorter
switching times, on a picosecond time scale,
in comparison with nanosecond magnetic control of the spin, 
which is essential for the implementation 
of quantum information processing schemes.
The capability of encoding and manipulating information at the single-spin level
is of great importance and has been experimentally demonstrated recently:
the generation \cite{dutt05} and optical control \cite{bayer06,dutt06}
of the spin coherence together with a possible
read-out of the state of a single confined spin in a QD system \cite{atature07}
can make the implementation of a quantum computer feasible.

A promising scheme of quantum optical control
of a spin in a single QD was recently proposed in Ref.~\onlinecite{chen04}.
It has been shown that coupling to a trion (charged exciton) state
leads to an arbitrary rotation between the two Zeeman-split spin states.
In this way, optical coherent control of a spin in a QD 
via adiabatic Raman transitions is possible.
This control protocol does not require an auxiliary fourth state
which was needed in a similar scheme previously proposed \cite{troiani03},
removing the requirement for the transfer of the electron
between two QDs and the delocalized hole state.

In realistic experiments, the QDs used for the proposed implementation are 
embedded in a solid state matrix, and confined carriers interact with the 
phonon bath, leading to a loss of coherence. In optical spin control 
schemes, spin rotation is achieved by spin-dependent charge evolution. 
Therefore, the spin decoherence in these schemes results mainly from the 
lattice response to the evolving charge density 
\cite{calarco03,roszak05,grodecka05,parodi06}. This effect has been studied 
\cite{roszak05} for a control scheme using an auxiliary state 
\cite{troiani03} but not for the scheme of Ref.~\onlinecite{chen04} 
that seems to be advantageous in some respects. 
Because of the different scheme of control the 
phonon-induced errors in the latter case have a different form and can be 
attributed to two channels: pure dephasing and phonon-assisted trion 
generation.
Furthermore, since in this control procedure the spin rotation involves a finite
trion occupation  
(unlike in the other scheme\cite{troiani03}), in addition to the
phonon interaction the  
finite trion lifetime can also lead to decoherence, because of the nonzero 
probability of the radiative decay of the trion \cite{caillet07}. Moreover, 
the operation in the ideal case should be performed adiabatically and the 
imperfect adiabaticity of the evolution will contribute to the total error of 
the operation. Finally, since this control scheme involves spectral selection 
of transitions (in contrast to the other one), off-resonant terms can also 
lead to an unwanted leakage to the trion state and result in large 
discrepancies in the desired spin rotation. It is not clear in advance which 
of these factors (if any) will dominate the decoherence under specific 
driving conditions.
So far, for the spin control scheme of Ref. \onlinecite{chen04}, only
the error resulting from imperfect adiabaticity of the driving and from the 
finite trion lifetime was evaluated \cite{caillet07} but the impact of 
phonons, off-resonant excitations, and the joint decoherence effect
have not been studied.

In this paper, we study the combined influence of the phonon
and photon environments
and of the imperfections of the evolution on an optical spin control scheme based on an
off-resonant coupling of the spin states to a trion state
in a doped semiconductor QD \cite{chen04}.
As a single-qubit gate, an optically induced arbitrary rotation
of the electron spin state is considered.
We show that the interactions with phonons and photons
are the dominant sources of decoherence in this system.
The phonon-induced decoherence has two origins:
pure dephasing and phonon-assisted trion generation,
with a different dependence on the duration and the detuning of the optical control pulses.
We show that by slowing down the evolution
one can considerably decrease only the error due to the pure dephasing,
while the phonon-assisted trion generation part still results in a large operation error.
The contributions to the error resulting from
the finite lifetime of the trion and the imperfections
of the evolution are also studied.
We calculate the total error of the operation and study the nontrivial
interplay of the different contributions.
In particular, we show that for moderate detunings ($\sim 1$~meV)
and long pulse durations the error is dominated by phonon-related effects.
We indicate the optimal conditions for the qubit rotation
for which the error is considerably small, with values even below $10^{-4}$,
which is essential for coherent quantum control,
and discuss the possible optimization against particular contributions to the error.

The paper is organized as follows.
In Sec.~\ref{sec:model}, we introduce the model for the qubit
based on the spin states of a confined electron in a charged QD.
Next, in Sec.~\ref{sec:rotation}, we describe a single-qubit rotation scheme.
In Sec.~\ref{sec:imperfections}, we study the imperfections of the evolution.
Section~\ref{sec:perturbation} describes the general perturbative method for describing the effects
of the environment on an arbitrary operation on a logical qubit.
Section~\ref{sec:decoherence} contains the results for the error
contributions due to the interaction of the carriers
with the phonon and photon environments.
In Sec.~\ref{sec:interplay}, we calculate the total error of the spin rotation
and discuss the interplay and possible optimization of the different sources of the error.
Section~\ref{sec:conclusion} concludes the paper with final remarks.
In addition, some technical details are presented in the Appendixes.

\section{Model system}
\label{sec:model}

\begin{figure}[b] 
\begin{center} 
\unitlength 1mm
{\resizebox{80mm}{!}{\includegraphics{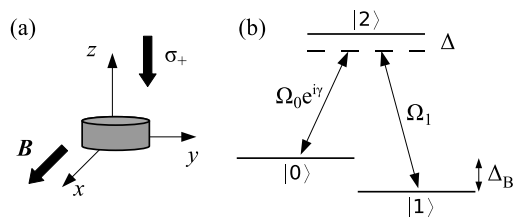}}}
\end{center} 
\caption{\label{fig:stirap} (a) Schematic plot of the QD with
the direction of incidence of the $\sigma_{+}$-polarized laser beam 
and the orientation of the magnetic field.
(b) Three-level system in a single doped QD.}
\end{figure}

We consider a single semiconductor QD charged with one additional electron.
A magnetic field applied in the $x$ direction 
[see Fig.~\ref{fig:stirap}(a)]
causes a Zeeman splitting
$\Delta_{\rm B}$ between the electron states with spin-up and spin-down
(with respect to the direction of the magnetic field)
which define the two logical qubit states $|0\rangle$ and $|1\rangle$
in the proposed scheme \cite{chen04}.
These two states are off-resonantly coupled (with a detuning $\Delta$)
to a trion state $|2\rangle$ by $\sigma_{+}$--polarized
laser pulses with real amplitudes $\Omega_{0}(t)$ and $\Omega_{1}(t)$
and with different phases [see Fig.~\ref{fig:stirap}(a,b)].
These two electron spin states $|0\rangle$ and $|1\rangle$
together with the trion state $|2\rangle$ compose
a three-level system, known in the literature as a $\Lambda$ system \cite{scully97}.

The Hamiltonian of the system is given by
\begin{equation*}
 H = H_{\mathrm{C}} + H_{\mathrm{env}} + V,
\end{equation*}
where the first term (control Hamiltonian) describes the carriers
and their interaction with the classical driving field (the laser
beam) and the second is the sum of the free phonon $H_{\mathrm{ph}}$
and photon $H_{\mathrm{rad}}$ contributions.
The last part, $V = H_{\mathrm{c-ph}}+ H_{\mathrm{c-rad}}$,
describes the coupling of the carriers to the environment,
where the first term denotes the carrier-phonon interaction
and the second describes the coupling of the carriers to the photon field.

The control Hamiltonian for this system in dipole
and rotating wave approximations is
\begin{eqnarray*}
H_{\mathrm C} & = & \sum_{n} \epsilon_{n} |n \rl n |  \\
&& + \frac{\hbar}{2}\left[ \sum_{n=0,1} \Omega_{n} (t) 
e^{i(\omega_{n} t + \gamma_{n})} 
(|0\rangle+|1\rangle)\langle 2| + \mathrm{H.c.} \right],
\end{eqnarray*}
where $\epsilon_{n}$ are the energies of the corresponding states, 
$\omega_{n}$ are the laser frequencies,
and $\gamma_{n}$ are the phases of the pulses.
The frequencies $\omega_{0}$ and $\omega_{1}$ have
to satisfy the Raman conditions, namely that the detunings
from the corresponding transition energies
$\epsilon_{2} - \epsilon_{n}$ should be the same. 
To this end, we set $\omega_{n} = (\epsilon_{2} - \epsilon_{n})/\hbar - \Delta$
($n = 0, 1$), where $\Delta$ is the common Raman detuning.
We perform a unitary transformation to the rotating frame with
$|\tilde n\rangle = e^{i(\omega_{n} t - \gamma_{0})} |n\rangle$ ($n=0,1$).
We can set $\gamma_{0} = \gamma$, $\gamma_{1}=0$ because only the
relative phase is important. In the rotating frame, the Hamiltonian reads
\begin{eqnarray}\label{hc}
 H_{\mathrm C} & = & \hbar \Delta |2 \rl 2 | + \frac{\hbar}{2} \Omega_{0} (t)
\left(e^{i \gamma} |\tilde 0\rl 2| + e^{-i \gamma}|2 \rl \tilde 0| \right) \\ \nonumber
&& + \frac{\hbar}{2} \Omega_{1} (t) 
\left(|\tilde 1\rl 2| + |2 \rl \tilde 1| \right)+H_{\mathrm{C}}',
\end{eqnarray}
where
\begin{equation*}
H_{\mathrm{C}}'=
\frac{\hbar}{2}\Omega_{1}(t) e^{i \Delta_{\rm B} t} |\tilde 0\rl 2|
+\frac{\hbar}{2}\Omega_{0}(t)e^{-i \Delta_{\rm B} t + i \gamma} |\tilde 1\rl 2| 
+ {\rm H.c.}
\end{equation*}
contains oscillating (off-resonant) terms which can be treated as a perturbation to
the ideal evolution generated by the Hamiltonian given in Eq.~(\ref{hc}).

The free phonon Hamiltonian has the form
\begin{equation*}
 H_{\mathrm{ph}} = \sumk\hbar\wk^{\phantom{\dag}} \bekd \bek^{\phantom{\dag}},
\end{equation*}
where $\bekd$ and $\bek$ are phonon creation and annihilation
operators, respectively, with corresponding frequencies $\wk$,
where $\kk$ is the phonon wave number.
The unperturbed photon Hamiltonian in the absence of charges reads
\begin{equation*}
 H_{\mathrm{rad}} = \sum_{{\bf q},\lambda}\hbar\omega'_{\bf q}
c_{{\bf q} \lambda}^{\dag} c_{{\bf q} \lambda}^{\phantom{\dag}}
\end{equation*}
with photon creation and annihilation operators
$c_{{\bf q} \lambda}^{\dag}$ and $c_{{\bf q} \lambda}$, respectively.
Here, $\omega'_{\bf q} = c |{\bf q}|/n_{\mathrm{r}}$
is the photon frequency, with the speed of light in vacuum $c$,
the photon wave number ${\bf q}$, and the refractive index of the semiconductor medium
$n_{\mathrm{r}}$; $\lambda$ labels the polarization. 
The unperturbed evolution of the system and of the decoupled environment
is described by $H_{\mathrm{C}}+H_{\mathrm{ph}}+H_{\mathrm{rad}}$.

The interaction of the carriers with the environment
includes the coupling to phonons and photons.
The carrier-phonon interaction reads
\begin{equation}\label{hcph}
 H_{\mathrm{c-ph}} = \sum_{n,n'} |n \rl n'| \sumk f_{nn'}(\kk)
 \left(\bek^{\phantom{\dag}} + \bemk^{\dagger} \right)
\end{equation}
with coupling constants $f_{nn'}(\kk)$ having the symmetry
$f_{nn'}(\kk) = f_{n'n}^{*}(-\kk)$.
The states $|0\rangle$ and $|1\rangle$ correspond to a single electron
confined in the same QD structure and differ only by the spin orientation.
Therefore, they have the same orbital wave functions, and thus the coupling constants 
$f_{00}(\kk)$ and $f_{11}(\kk)$ are the same.
These states have different spins so that $f_{01}(\kk)$ and
$f_{10}(\kk)$ would describe a `direct' phonon-assisted spin flip,
mediated by the spin-orbit coupling \cite{golovach04}. However, in the optical
spin-control schemes, the effect of this process is many orders of
magnitude weaker than the decoherence due to the dynamical response to
charge evolution \cite{roszak05}. Therefore, we neglect this coupling
and set these coefficients equal to zero. 

Initially, before the arrival of the pulses, the lattice
is in a configuration where one electron is present in the QD
surrounded by a lattice deformation
(a polaron-like state \cite{vagov02,jacak03,machnikowski07}).
We redefine the phonon modes to obtain the ground state
of the carrier-phonon system corresponding to this new lattice equilibrium.
In terms of the new phonon operators
\begin{equation*}
 \bk = \bek + \frac{f_{00}(\kk)}{\hbar\wk},
\end{equation*}
the interaction with the lattice modes reads
\begin{eqnarray}\label{Hc-ph}
 H_{\mathrm{c-ph}} & = & |2 \rl 2| \sumk F_{22}(\kk)\left(\bk^{\phantom{\dag}}+\bmkd \right) \\ \nonumber
 && + \left[ | \tilde 1 \rl 2| \sumk F_{12}(\kk)\left(\bk^{\phantom{\dag}}+\bmkd \right) + \mathrm{H.c.} \right],
\end{eqnarray}
where $F_{22}(\kk) = f_{22}(\kk) - f_{00}(\kk)$ 
and $F_{12}(\kk) = f_{12}(\kk) e^{-i(\omega_{1}t-\gamma_{0})}$.
Additionally, there is a polaron-like energy shift
which is included in the control Hamiltonian $H_{\mathrm{C}}$.
The interband off-diagonal phonon coupling [the second term in Eq.~(\ref{Hc-ph})]
has a negligible effect due to energetic reasons \cite{roszak05} and can be disregarded.
For spectrally narrow pulses, which are needed for the adiabaticity of
the procedure, only acoustic phonons contribute to the dephasing. For
overlapping electron and hole wave functions, the excitation of a
confined trion does not involve considerable charge redistribution 
and the effect of the piezoelectric coupling is very weak \cite{krummheuer02}.
Therefore, we consider only the interaction with longitudinal acoustic phonons
via the deformation potential coupling.

We assume for simplicity that the trion state is described
by a product of electron and hole wave functions 
$\Psi_{\rm e(h)}(\bm{r})$ which are the same as those corresponding to
a single confined carrier. This is a reasonable
approximation in the strong confinement limit,
where the Coulomb interaction is of minor influence on the wave functions
and leads only to energy renormalization effects which are included in
the transition energies. 
The coupling between the trion and phonons 
[the first term in Eq.~(\ref{Hc-ph})] has the form (see Appendix \ref{app:couplings})
\begin{equation}\label{f22}
F_{22}(\kk) = f_{22}(\kk)-f_{00}(\kk)=
\sqrt{\frac{\hbar k}{2 \rho V c_{\mathrm l}}}
(D_{\mathrm e}  -D_{\mathrm h}) \mathcal{F}(\kk),
\end{equation}
where $\rho$ is the crystal density,
$V$ is the normalization volume of the phonon modes,
$c_{\mathrm{l}}$ is the longitudinal speed of sound, and $D_{\mathrm{e(h)}}$
is the deformation potential constant for the electron (hole).
The form factor $\mathcal F(\kk)$
depends on the geometry of the wave functions
$\Psi_{\rm e(h)}(\bm{r})$. We assume Gaussian wave functions
\begin{equation}\label{wavefunction}
 \Psi_{\rm e(h)}(\bm{r}) \sim \exp{\left(-\frac{r_{\perp}^{2}}{
2l_{\rm e(h)}^{2}} -\frac{z^{2}}{2 l_{z}^{2}} \right)},
\end{equation}
where $l_{\mathrm{e(h)}}$ is the confinement size for an electron (a hole) in
the $xy$ plane, $l_{z}$ is the common confinement size along $z$, and
$r_{\bot}$ and $r_{z}$ are the corresponding components of the position.
Then, neglecting the small correction resulting from the difference
between the electron and hole confinement sizes, one finds
\begin{equation}\label{formfactor}
\mathcal{F}(\kk) = e^{-(k_{\bot}l/2)^{2}-(k_{z}l_{z}/2)^{2}},
\end{equation}
where $l^{2}=(l_{\mathrm{e}}^{2}+l_{\mathrm{h}}^{2})/2$ and
$k_{\bot}$ and $k_{z}$ are the components of the wave vector in the
$xy$ plane and along $z$, respectively (see Appendix \ref{app:couplings} for details).

The carrier-photon interaction Hamiltonian in the rotating wave
approximation reads (in the rotating frame)
\begin{eqnarray}\label{hcrad}
H_{\rm{c-rad}} & = & \frac{1}{\sqrt{2}}\sum_{{\bf q},\lambda}
g_{{\bf q} \lambda}^{\phantom{\dag}} c_{{\bf q} \lambda}^{\dag}
\left[ e^{i \omega_{0} t} |\tilde{0}\rl 2|
\right. \\ \nonumber && \left. + e^{i (\omega_{1} t - \gamma)}
 |\tilde{1}\rl 2| \right] + \mathrm{H.c.}
\end{eqnarray}
with the coupling constants
\begin{equation*}
g_{{\bf q} \lambda} = -i \sum_{ \alpha = 1}^{3} d_{ \alpha }
\sqrt{\frac{\hbar \omega'_{\bf q}}{2 \epsilon_{0} \epsilon_{\rm r} V}}
e_{ \alpha }^{(\lambda)}({\bf q}).
\end{equation*}
Here, $\alpha$ denotes Cartesian components,
$d_{ \alpha }$ is the interband dipole moment, 
$\epsilon_{0}$ and $\epsilon_{\rm r}=n_{\mathrm{r}}^{2}$ are the
vacuum dielectric constant 
and the semiconductor relative dielectric constant, respectively,
and $\bm{e}^{(\lambda)}({\bf q})$ is the unit polarization vector.

In Tab.~\ref{tab:param}, the material parameters (corresponding to a
self-assembled InAs/GaAs system) are given.

\begin{table}
\begin{tabular}{lll}
\hline
Deformation potential coupling 
	& $D_{\mathrm{e}}-D_{\mathrm{h}}$ & 8 eV \\
Crystal density & $\rho$ & 5360 kg/m$^3$ \\
Speed of sound (longitudinal) & $c_{\mathrm{l}}$ & 5150 m/s \\
Wave function width in-plane & $l$ & 5 nm \\
Wave function width in $z$ direction & $l_{z}$ & 1 nm \\
Trion decay rate & $\Gamma$ & 1 ns$^{-1}$ \\
Zeeman splitting & $\hbar\Delta_{\rm B}$ & 1 meV \\
\hline
\end{tabular}
\caption{\label{tab:param}System parameters used in the calculations.}
\end{table}


\section{Unperturbed spin rotation}
\label{sec:rotation}

In this section, we present the formal description
of the spin rotation procedure \cite{chen04}
without the interaction with the environment.
In the ideal case, the evolution is slow and may
be described by invoking the adiabatic theorem \cite{messiah66}.

In the considered three-level system [Fig.~\ref{fig:stirap}(b)],
it is possible to perform an arbitrary rotation
of the electron spin via the intermediate trion state $|2\rangle$.
To show this, one sets in the control Hamiltonian $H_{\mathrm{C}}$ [Eq.~(\ref{hc})]
\begin{displaymath}
\Delta  =  \Theta(t) \cos[2\phi(t)]
\end{displaymath}
and 
\begin{displaymath}
\Omega_{0}(t) = \Omega(t)\cos\beta,\quad
\Omega_{1}(t) = \Omega(t)\sin\beta,
\end{displaymath}
where 
\begin{displaymath}
\Omega(t) = \Theta(t) \sin[2\phi(t)].
\end{displaymath}
Here, $\Theta(t)=\sqrt{\Omega^{2}(t)+\Delta^{2}}$
is the (time-dependent) effective Rabi frequency and 
\begin{equation*}
\sin^{2}\phi(t) = \frac{1}{2}\left( 
1-\frac{\Delta}{\sqrt{\Omega^{2}(t)+\Delta^{2}}} \right).
\end{equation*}
For $\Delta>0$, switching the pulses off corresponds to $\phi\to 0$.

To ensure an adiabatic evolution, the angle $\beta$,
defined via $\tan \beta = \Omega_{1}/\Omega_{0}$,
should vary slowly in time. We choose $\Omega_{0}$ and $\Omega_{1}$ to
have the same envelope shapes, so that $\beta$ becomes time independent.

We introduce new basis states
\begin{eqnarray*}
|B\rangle & = & e^{i\gamma} \cos\beta |\tilde 0\rangle + \sin\beta |\tilde 1\rangle,\\
|D\rangle & = & e^{i\gamma} \sin\beta |\tilde 0\rangle - \cos\beta |\tilde 1\rangle,
\end{eqnarray*}
which are superpositions of the qubit states
$|0\rangle$ and $|1\rangle$ selected by the laser pulses,
where only the \textit{bright} state $|B\rangle$
is coupled to the trion state and the orthogonal \textit{dark} state
$|D\rangle$ remains unaffected.
The Hamiltonian $H_{\mathrm{C}}$ which generates the ideal evolution
 [Eq.~(\ref{hc})] now reads
\begin{equation*}
H_{\rm C} = \hbar\Delta |2\rl 2| + \frac{\hbar}{2}\Omega(t)
\left( |B \rl 2| + |2 \rl B| \right)
\end{equation*}
and has the instantaneous eigenstates
\begin{subequations}
\begin{eqnarray}\label{basis}
|a_{0}(t)\rangle & = & |D\rangle, \\
|a_{1}(t)\rangle & = & \cos\phi(t) |B\rangle -\sin\phi(t) |2\rangle,\\
|a_{2}(t)\rangle & = & \sin\phi(t) |B\rangle + \cos\phi(t) |2\rangle
\end{eqnarray}
\end{subequations}
with the corresponding eigenvalues
\begin{eqnarray*}
\lambda_{0}(t) & = & 0,\\ 
\lambda_{1}(t) & = & -\hbar\Theta(t)\sin^{2}\phi(t)
= \frac{\hbar}{2}\left(\Delta - \sqrt{\Delta^2 + \Omega^{2}(t)} \right),\\
\lambda_{2}(t) & = & \hbar\Theta(t)\cos^{2}\phi(t)
= \frac{\hbar}{2}\left(\Delta + \sqrt{\Delta^2 + \Omega^{2}(t)} \right). 
\end{eqnarray*}

\begin{figure}[tb]
\begin{center} 
\unitlength 1mm
{\resizebox{85mm}{!}{\includegraphics{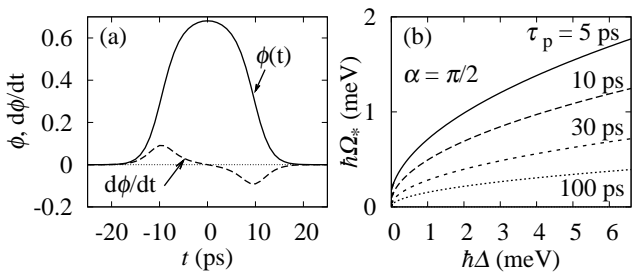}}}
\end{center} 
\caption{\label{fig:omega} (a) The tipping angle and its derivative
as a function of time for a detuning $\hbar\Delta = 1$~meV and a
pulse duration $\tau_p=5$~ps.
(b) The adjusted amplitude of the control pulse
for the $\pi/2$ rotation about the $z$ axis
as a function of detuning for different fixed pulse durations.}
\end{figure}

The system evolution is realized by the change of the so called
tipping angle $\phi(t)$ [see Fig.~\ref{fig:omega}(a)], i.e.,
by the change of the pulse amplitudes.
The condition for adiabaticity is a slow change of $\phi(t)$
in comparison with the rate of the adiabatic motion 
given by the effective Rabi frequency, $|\dot{\phi(t)}|\ll \Theta(t)$.
If the adiabatic condition is met, the state of the system
(initially a combination of $|0\rangle$ and $|1\rangle$)
remains in the subspace spanned by the two eigenstates $|a_{0}\rangle$
and $|a_{1}\rangle$ during the whole process.

The evolution operator $U_{\mathrm C}(t)$ 
in the absence of the environment perturbation
(in the basis $|D\rangle$, $|B\rangle$, $|2\rangle$) has the form
\begin{equation}\label{Uan}
U_{\mathrm{C}} (t) =
    \left( \begin{array}{ccc}
    1 & 0 & 0 \\
    0 & e^{-i\Lambda_{1}(t)} \cos\phi(t) & e^{-i\Lambda_{2}(t)} \sin\phi(t) \\
    0 & -e^{-i\Lambda_{1}(t)} \sin\phi(t) & e^{-i\Lambda_{2}(t)} \cos\phi(t)
    \end{array}\right),
\end{equation}
where
\begin{equation*}
\Lambda_{n}(t) = \frac{1}{\hbar}
\int_{t_{0}}^{t} d\tau \lambda_{n}(\tau), \;\;\; (n = 1,2).
\end{equation*}
From Eq.~(\ref{Uan}), it is clear that
after the operation, when the pulses are switched off ($\phi\to 0$),
an arbitrary initial electron spin state
will have acquired a phase in the $|B\rangle$ component
with respect to the orthogonal dark superposition $|D\rangle$.
The resulting unitary transformation in the qubit subspace can be
written as 
\begin{equation*}
U_{\mathrm{C}}(\infty) = e^{-\frac{i}{2}\Lambda_{1}\vec \sigma \cdot \vec n} =
\cos\frac{\Lambda_{1}}{2}\mathbf{I} -i \sin\frac{\Lambda_{1}}{2} \vec
\sigma \cdot \vec n,
\end{equation*}
where $\mathbf{I}$ is the unit operator and $\vec{\sigma}$ is the
vector of Pauli matrices
in the original qubit basis $|0\rangle$, $ |1\rangle$. 
This transformation corresponds to a rotation
through an angle $\Lambda_{1}(\infty)$ about the axis $\vec n$,
\begin{equation*}
\vec n = [ -\cos \gamma \sin(2 \beta), \sin \gamma\sin(2\beta), -\cos(2\beta)],
\end{equation*}
which depends on the ratio of the pulse envelopes
and on the relative phase of the pulses.

For convenience, we write the general initial state of the spin qubit
in the form 
\begin{equation}\label{psi0}
|\psi_{0}\rangle = \cos \frac{\vartheta}{2}|B\rangle 
+e^{i\varphi} \sin\frac{\vartheta}{2} |D\rangle,
\end{equation}
where $\vartheta$ and $\varphi$ are angles on a Bloch sphere.
We will also need two states orthogonal to $|\psi_{0}\rangle$, which we
choose in the form
\begin{equation*}
|\psi_{1}\rangle = \sin \frac{\vartheta}{2}|B\rangle
- e^{i\varphi} \cos\frac{\vartheta}{2} |D\rangle,
\;\;\;\;\;\;\; |\psi_{2}\rangle = |2\rangle.
\end{equation*}

We assume that the rotation is performed using Gaussian control pulses
\begin{equation*}
\Omega(t) = \Omega_{*} \exp{\left(-\frac{t^{2}}{2\tau_{\rm
p}^{2}}\right)}, 
\end{equation*}
where $\Omega_{*}$ is the amplitude of the control pulse
and $\tau_{\rm p}$ its duration.
In the following discussion, we will treat the 
pulse duration $\tau_{\rm p}$ and the detuning $\Delta$ as tunable
parameters, while the pulse amplitude $\Omega_{*}$ will be adjusted
to achieve the desired rotation of the qubit.
The amplitude as a function of detuning
for a $\pi/2$ rotation about the $z$ axis [growth direction,
see Fig.~\ref{fig:stirap}(a)]
is plotted for different pulse durations in Fig.~\ref{fig:omega}(b).

\section{Imperfections of the unitary evolution}
\label{sec:imperfections}

Before we study the impact of the environment, let us discuss the
limitations imposed on the driving by the requirement of adiabatic
evolution, as well as the effect of the oscillatory terms contained in
$H_{\mathrm{C}}'$ in Eq.~(\ref{hc}) and neglected in the discussion
presented in the previous section (the former has also been studied in
Ref. \onlinecite{caillet07}).

A perfectly adiabatic evolution does not transfer the qubit states
$|0\rangle$ and $|1\rangle$ into the trion state $|2\rangle$,
which is only slightly occupied during the gating.
In realistic experiments, the parameters cannot be changed
infinitely slowly, so that there is a nonzero probability of a jump from
the ideal instantaneous (adiabatic) state to one of the other states.
Representing the exact system state in terms of the adiabatic
eigenstates [Eqs.~(\ref{basis}-c)], 
\begin{equation*}
|\psi\rangle=\sum_{n}c_{n}(t)e^{-i\Lambda_{n}(t)}|a_{n}(t)\rangle,
\end{equation*}
one finds the equation for the probability amplitudes \cite{messiah66},
\begin{equation*}
 \dot{c}_{m}(t) = -\sum_{n}e^{i\left[\Lambda_{m}(t)-\Lambda_{n}(t)\right]}
\langle a_{m}(t) | \dot{a}_{n}(t) \rangle c_{n}(t).
\end{equation*}
The state $|a_{0}\rangle$ is time independent
and represents the dark state $|D\rangle$ decoupled from the trion state $|2\rangle$.
Moreover, $\langle a_{0}(t)|\dot{a}_{1}(t)\rangle=0$. Therefore, to the leading
order, the only unwanted transition is to the state $|a_{2}\rangle$,
which becomes the trion state after switching off the pulses.
The corresponding amplitude is
\begin{equation*}
 c_{2}^{\mathrm{na}}\approx -\int_{-\infty}^{\infty} dt \;
e^{i\left[\Lambda_{2}(t)-\Lambda_{1}(t)\right]}
\langle a_{2}(t) | \dot{a}_{1}(t) \rangle c_{1}(0).
\end{equation*}
If the initial state is $|\varphi_{0}(t)\rangle$, as given by
Eq.~(\ref{psi0}), then $c_{1}(0)=\cos(\vartheta/2)$ and
\begin{equation}\label{eq:cna}
c_{2}^{\mathrm{na}}\approx \cos \frac{\vartheta}{2}
\int_{-\infty}^{\infty} dt \; e^{i\left[\Lambda_{2}(t)-\Lambda_{1}(t)\right]} \dot{\phi}(t).
\end{equation}

The other source of imperfections in the controlled evolution are the
rotating (off-resonant) terms contained in the Hamiltonian
$H_{\mathrm{C}}'$, which reads in the new basis
\begin{eqnarray*}
\lefteqn{H_{\mathrm{C}}' = } \\
&& \frac{\hbar}{2}\Omega(t) \left[
e^{i(\Delta_{\rm B} t - \gamma)} \sin^{2}\beta
- e^{-i(\Delta_{\rm B} t - \gamma)} \cos^{2}\beta \right] |D \rl 2| \\
&& + \frac{\hbar}{2} \Omega(t) \cos(\Delta_{\rm B} t - \gamma)
\sin(2 \beta) |B \rl 2| + {\rm H.c.}
\end{eqnarray*}
We assume that the correction to the unitary evolution resulting from
these terms is small and treat them perturbatively. 
The effects of the additional Hamiltonian $H'_{\mathrm{C}}$ may be of
two kinds: additional unitary
rotation within the computational space and leakage to the
trion state. The former can be taken into account when
designing the control pulses and compensated by a suitable modification
of the control parameters. Therefore, we treat only the latter as an
error. The amplitude for the trion excitation is given by 
\begin{widetext}
\begin{eqnarray*}
c_{2}^{\mathrm{off}} & = & \int_{-\infty}^{\infty}dt
\langle 2|H_{\mathrm{C}}'(t) |\psi_{0}\rangle = \\ &&
\frac{1}{2} \cos \frac{\vartheta}{2} \sin(2\beta)
\int_{-\infty}^{\infty}dt \; e^{i\left[\Lambda_{2}(t)-\Lambda_{1}(t)\right]} \Omega (t) 
\cos (\Delta_{\rm B} t - \gamma) \cos [2 \phi(t)] \\
&& + \frac{1}{2} e^{i \varphi} \sin \frac{\vartheta}{2}
\int_{-\infty}^{\infty}dt \; e^{i \Lambda_{2}(t)}\Omega (t) \cos\phi(t)
\left[ e^{-i(\Delta_{\rm B}t - \gamma)} \sin^{2}\beta
- e^{i(\Delta_{\rm B}t - \gamma)} \cos^{2}\beta \right],
\end{eqnarray*} 
\end{widetext}
where 
$H_{\mathrm{C}}'(t)=U^{\dag}_{\rm C}(t)H_{\mathrm{C}}'U^{\phantom{\dag}}_{\rm C}(t)$
is the relevant Hamiltonian in the interaction picture 
with respect to the perfect evolution described by the evolution operator
in Eq.~(\ref{Uan}).

\begin{figure}[tb]
\begin{center} 
\unitlength 1mm
{\resizebox{85mm}{!}{\includegraphics{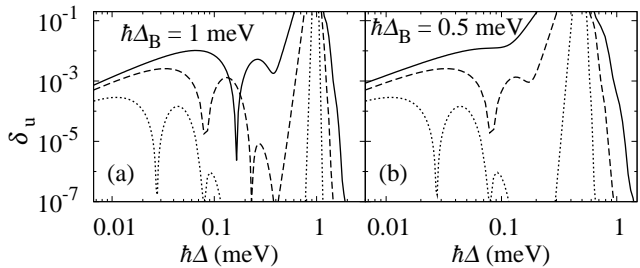}}}
\end{center} 
\caption{\label{fig:error-u} The error due to the unitary corrections
as a function of detuning for two values 
of the Zeeman splitting $\Delta_{\rm B}$ and
three values of the pulse duration: $\tau_{\rm p} = 5$~ps (solid lines),
$\tau_{\rm p} = 10$~ps (dashed lines), and $\tau_{\rm p} = 30$~ps (dotted lines).}
\end{figure}

The error due to both unitary corrections described above is
given by the total probability of leakage to the trion state,
\begin{equation*}
\delta_{\mathrm{u}}
=\left|c_{2}^{\mathrm{na}}+c_{2}^{\mathrm{off}}\right|^{2},
\end{equation*}
and is plotted in Fig.~\ref{fig:error-u}
for a $\pi/2$ rotation about the $z$ axis
as a function of detuning for two values of the Zeeman splitting
and different pulse durations.
If the values of the detuning approach the Zeeman splitting $\Delta_{\rm B}$,
the error becomes very large, since one spin state is
then almost resonantly coupled to the trion state. 
This induces a large trion occupation
which inhibits a coherent spin rotation in this scheme.
If the evolution is fast (short pulse durations $\tau_{\rm p}$),
the adiabatic condition is not met for small detunings, which results
in larger errors.

In addition, the unitary error shows oscillations, 
visible in Figs.~\ref{fig:error-u}(a,b),
which are due to the nonadiabatic contribution. In order to understand their
origin, let us note that, in spite of the smooth Gaussian pulse envelope, the
tipping angle evolves in a step-wise manner [see Fig.~\ref{fig:omega}(a)]
(especially for strong pulses).
Thus, $\dot{\phi}(t)$ has two peaks of opposite signs at $t=\pm t_1$, where
$t_1$ is a certain time, depending on the pulse duration $\tau_{\rm p}$
($t_1\approx 10$~ps in Fig.~\ref{fig:omega}(a)]). Hence, according 
to Eq.~(\ref{eq:cna}), the probability for a
non-adiabatic jump is approximately proportional to
$\sin^{2}[\Lambda_{2}(t_1)-\Lambda_{1}(t_1)-\Lambda_{2}(-t_1)+\Lambda_{1}(-t_1)]$
and is therefore an oscillating function of the control parameters.
This oscillation  of the transition probability reflects the interference of
the amplitudes for non-adiabatic jumps when the trion is switched on and off.

For reasonably long pulses, one can indicate two regimes of the detuning
where the operation on the qubit has a high fidelity.
The values of the detuning have to be chosen either above the Zeeman splitting
or below it and larger than $0.1$~meV.
To take advantage of the available fast optical control methods
and insure the adiabaticity of the evolution,
one can perform the operation with short pulse durations of several picoseconds
and detunings of a few~meV.


\section{Environment perturbation during operation on the qubit}
\label{sec:perturbation}

In this section, we summarize the general method 
for describing the effects of a perturbation due to the environment
on an arbitrary operation on a qubit. A full account of this  approach
is given in earlier works\cite{grodecka05,roszak05}. Here, we give a
brief overview for the sake of completeness and for reference in the
following sections.

The evolution in the absence of the perturbation due to the environment
is described by the unitary evolution operator
$U_{0}(t) = U_{\mathrm C}(t) \otimes e^{-i H_{\mathrm{env}} t}$.
The effect of the interaction with the environment is calculated using
the second-order Born expansion of the evolution equation for the density matrix.
We include the effect of the driving field non-perturbatively and
treat the carrier-environment interaction within a perturbation theory. 

The reduced density matrix of the qubit reads
\begin{equation}
\label{dens-mat}
 \rho (t) = U_{0}(t) [\rho_{0}+\rho^{(2)}(t)] U_{0}^{\dag}(t),
\end{equation}
where $\rho^{(2)}(t)$ is the correction to the density matrix
resulting from the interaction with the environment (in the
interaction picture)
and $\rho_{0}$ the initial state of the qubit,
which is assumed to be pure, $\rho_{0} = |\psi_{0} \rl \psi_{0} |$.
The initial state of the system (qubit together with the environment)
has the form $\rho_{0}\otimes \rho_{T}$, where $\rho_{T}$
is the thermal equilibrium state of the environment bath.

To quantify the quality of the operation on a qubit,
we use the \textit{fidelity}
\cite{nielsen00} 
\begin{equation}
\label{fidel}
F = \langle \psi_{0} |U_{0}^{\dag}(\infty)  \rho (\infty) U_{0}(\infty) |
\psi_{0} \rangle^{1/2},
\end{equation}
which is a measure of the overlap between the ideal (pure) final 
state without perturbation, $U_{0}(\infty)|\psi_{0}\rangle$, and the actual final
state of the system given by the density matrix $\rho (\infty)$.
If the procedure is performed ideally, i.e. without discrepancies
from the desired qubit operation, then $F=1$.
The fidelity loss  $\delta = 1 - F^{2}$ is referred to 
as the \textit{error} of the quantum gate. Inserting
Eq.~(\ref{dens-mat}) into Eq.~(\ref{fidel}), with 
$\rho_{0}=|\psi_{0}\rl\psi_{0}|$, one finds
\begin{equation}\label{error}
\delta = -\langle \psi_{0} | \rho^{(2)} (\infty) | \psi_{0} \rangle.
\end{equation}

The density matrix correction is calculated from a perturbation expansion
\cite{cohen98}:
\begin{equation*}
 \rho^{(2)}(t) = 
-\frac{1}{\hbar^{2}}\int_{t_{0}}^{t}d\tau\int_{t_{0}}^{\tau}d\tau'
      \tr_{\mathrm{R}}[V(\tau),[V(\tau'),\rho_{0}]],
\end{equation*}
where $\tr_{\rm R}$ is the trace with respect to the reservoir degrees
of freedom 
and $V(t) = U_{0}^{\dag}(t) V U_{0}(t)$ is the carrier-environment
Hamiltonian in the interaction picture.
It can always be written in the general form
\begin{equation*}
V = \sum_{nn'} S_{nn'} \otimes R_{nn'},
\end{equation*}
where $S_{nn'}$ acts on the carrier subsystem, 
$R_{nn'}$ acts on the environment of phonons or photons,
and $R_{nn'}=R_{n'n}^{\dag}$, $S_{nn'}=S_{n'n}^{\dag}$. 
It is easy to see that Eqs.~(\ref{hcph}) and (\ref{hcrad}) have this structure.

It is convenient to introduce two sets of spectral functions. 
The first is a family of spectral densities of the reservoir (phonons or
photons), defined as 
\begin{equation}\label{spectral}
 R_{nn',mm'}(\omega) 
= \frac{1}{2\pi}\int dt \; e^{i \omega t} \langle R_{nn'}(t) R_{mm'} \rangle,
\end{equation}
with the operator $R_{nn'}$ transformed into the interaction picture
$R_{nn'}(t) = U^{\dag}_{0}(t) R_{nn'} U_{0}(t)$.
The functions from the second set are nonlinear spectral characteristics
of the driven evolution,
\begin{equation}
\label{S}
S_{nn',mm'}(\omega) = 
\sum_{i}  \langle \psi_{0} | Y_{n'n}^{\dag}(\omega) | 
\psi_{i}\rl\psi_{i}|Y_{mm'}|\psi_{0}\rangle,
\end{equation}
where $|\psi_{i}\rangle$ span the subspace orthogonal to the initial state
$|\psi_{0}\rangle$ and 
\begin{equation}\label{Y}
 Y_{nn'}(\omega) =  \int dt \; S_{nn'}(t) e^{-i \omega t},
\end{equation}
with $S_{nn'}(t) = U^{\dag}_{0}(t) S_{nn'} U_{0}(t)$.
The various terms in Eq.~(\ref{S}) describe transitions to different
states orthogonal to the desired state $|\psi_{0}\rangle$. In the long
time limit for a time-independent system, all of them either vanish or
turn into energy-conserving Dirac delta functions, restoring Fermi's golden
rule for transition probabilities \cite{alicki02}. In the general case, they are
broadened due to time dependence.

Using the definitions in Eqs.~(\ref{spectral}) and (\ref{S}),
the error [Eq.~(\ref{error})] can be written in the form
\cite{grodecka05,roszak05} 
\begin{equation}\label{error-w}
\delta = \sum_{nn',mm'}\int d\omega \; R_{nn',mm'}(\omega) S_{nn',mm'}(\omega).
\end{equation}
The error can thus be expressed as an overlap between spectral functions,
which are the ``building blocks'' for the calculation
of the environment effects on the quantum evolution.
A detailed derivation of 
Eq. (\ref{error-w}) can be found in Refs. \onlinecite{roszak05} and
\onlinecite{grodecka05}. 

The perturbative approach described above obviously yields only an
approximate description of decoherence. In the case of phonon-induced
dephasing, comparisons with exact results
(for ultrashort laser pulses) \cite{axt05} and
with correlation expansion results \cite{krugel05} show that the
perturbative results are very accurate as long as the overall
dephasing effect is weak: the inaccuracy is of the order of
$\delta^{2}$, where $\delta$ is some measure of decoherence, e.g., the
fidelity loss, Eq.~(\ref{error-w}). The same holds for the
decoherence induced by the radiative decay of the trion, as we show in
Appendix \ref{app:lindblad}.

Thus, to calculate the error of the quantum gate
due to the interaction with the environment,
one needs to derive the two spectral functions.
These will yield a transparent spectral interpretation
for the various contributions to the qubit dephasing
and provide a possibility to seek the optimal conditions
depending on the system properties.


\section{Decoherence mechanisms}
\label{sec:decoherence}

In this section, the different kinds of decoherence mechanisms
(the coupling of the carriers to phonons and photons) are studied.
We apply the general theory introduced in the former section
to calculate the total error of a spin rotation through
an angle of $\pi/2$ about the $z$ axis.
The quantitative estimates are given 
for charged self-assembled InAs/GaAs quantum dot.

\subsection{Interaction with the phonons}

It follows from Eq.~(\ref{Hc-ph}) that the carrier-phonon interaction
is described by just one pair of operators,
$S_{22}=|2\rl 2|$ and $R_{22}=\sumk F_{22}(\kk)(\bk^{\phantom{\dag}}+\bmkd)$.
Therefore, one needs only two spectral functions:
the spectral density of the phonon reservoir 
$R_{22,22}(\omega)\equiv R_{\rm ph}(\omega)$
and the spectral characteristics of the driving 
$S_{22,22}(\omega)\equiv S_{\rm ph}(\omega)$
to calculate the phonon-induced error
\begin{equation}\label{error-ph}
 \delta_{\rm ph} = \int d\omega R_{\rm ph}(\omega) S_{\rm ph}(\omega).
\end{equation}
Using Eq.~(\ref{spectral}), one finds the explicit form of the former:
\begin{eqnarray}\label{Rph}
 \lefteqn{R_{\rm ph}(\omega) = }  \\ \nonumber
 && \frac{1}{\hbar^{2}} \left[n_{\rm B}(\omega) + 1\right] \sumk |F_{22}(\kk)|^{2}
\left[  \delta(\omega-\wk) + \delta(\omega+\wk) \right],
\end{eqnarray}
where $n_{\rm B}(\omega)$ is the Bose distribution function.

\begin{figure}[tb]
\begin{center} 
\unitlength 1mm
{\resizebox{85mm}{!}{\includegraphics{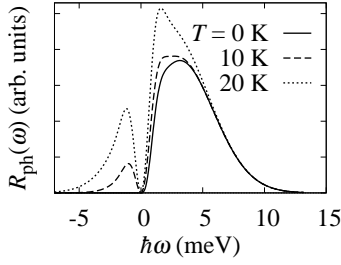}}}
\end{center} 
\caption{\label{fig:spectral-phonons} Spectral density of the phonon reservoir 
at different temperatures $T$.}
\end{figure}

The function $R_{\rm ph}(\omega)$
(see Appendix~\ref{app:couplings} for a detailed derivation)
is plotted in Fig.~\ref{fig:spectral-phonons} for different temperatures $T$.
The negative frequency part of the function corresponds
to phonon absorption by the carriers and is nonzero
only for finite temperatures, while the positive part
represents the emission processes and always has finite values.
The phonon spectral density has a cut-off at the frequency
$\omega \approx c_{\rm l}/l$,
which corresponds to the inverse of the time phonons need to traverse
the quantum dot. 

The carrier part of the interaction Hamiltonian
$S_{\mathrm{ph}} = |2\rl 2|$ leads to a spectral characteristics
containing two parts corresponding to the two orthogonal states:
\begin{eqnarray}
\label{sph}
\lefteqn{S_{\mathrm{ph}}(\omega)  = 
s_{1}^{\mathrm{ph}}(\omega) + s_{2}^{\mathrm{ph}}(\omega)} \\
\nonumber
&& = \frac{1}{4} \sin^{2}\vartheta \left| 
\int dt \; e^{i\omega t} \sin^{2} \phi(t) \right|^{2} \\ 
\nonumber
&& \phantom{11} + \frac{1}{4} \cos^{2}\frac{\vartheta}{2} \left| \int dt \; e^{i\omega t} 
e^{-i\left[\Lambda_{2}(t)-\Lambda_{1}(t)\right]} \sin \left[2 \phi(t)\right] \right|^{2}. 
\end{eqnarray}
The initial state, Eq.~(\ref{psi0}), can be an arbitrary superposition
of the states $|0\rangle$ and $|1\rangle$,
and the error depends on the choice of the initial qubit state.
In order to obtain representative error evaluations,
we average the error (thus, the contributing spectral function)
over the angles $(\vartheta, \varphi)$
on the Bloch sphere of the initial states, according to 
\begin{equation*}
s_{i\mathrm{(av)}}(\omega)=\frac{1}{4\pi}
\int_{0}^{\pi} d\vartheta\sin\vartheta
\int_{0}^{2\pi} d\varphi \; s_{i}(\omega).
\end{equation*}  
The averaged contributions to the spectral characteristics,
corresponding to the two terms in Eq.~(\ref{sph}), read
\begin{eqnarray*}
s_{1\mathrm{(av)}}^{\mathrm{ph}}(\omega) & = & 
\frac{1}{6} \left| \int dt \; e^{i\omega t} \sin^{2}\phi(t)\right|^{2},\\ \nonumber
s_{2\mathrm{(av)}}^{\mathrm{ph}}(\omega) & = & \frac{1}{8} 
\left|\int dt \; e^{i\omega t} e^{-i\left[\Lambda_{2}(t)-\Lambda_{1}(t)\right]}
\sin\left[ 2 \phi(t) \right] \right|^{2}.
\end{eqnarray*}
Both contributions to the phonon-induced error are independent of $\beta$. 

We can derive approximate analytical formulas for the spectral characteristics
under the condition that the control pulses are much smaller than the detuning
($\Omega \ll \Delta$), which is met for detunings 
of several meV [see Fig.~\ref{fig:omega}(b)].
Thus, in this detuning regime, we obtain
\begin{subequations}
\begin{eqnarray*}
s_{1}^{\mathrm{ph}}(\omega) & \approx &
\frac{\pi}{96} \frac{\Omega_{*}^{4} \tau_{\rm p}^{2}}{\Delta^{4}}
\exp{\left(-\frac{1}{2} \tau_{\rm p}^{2} \omega^{2} \right)},\\
s_{2}^{\mathrm{ph}}(\omega) & \approx & 
\frac{\pi}{4} \frac{\Omega_{*}^{2} \tau_{\rm p}^{2}}{\Delta^{2}}
\left\{ \exp{\left[-\frac{1}{2} \tau_{\rm p}^{2} (\Delta +\omega)^{2} \right]}\right.\\ \nonumber
&& \left. \phantom{aaaaaaa} -\frac{\Omega_{*}^{2}}{2\sqrt{3}\Delta^{2}}
\exp{\left[-\frac{1}{6} \tau_{\rm p}^{2} (\Delta + \omega)^{2} \right]} \right\}^{2}.
\end{eqnarray*}
\end{subequations}

\begin{figure}[tb] 
\begin{center} 
\unitlength 1mm
{\resizebox{85mm}{!}{\includegraphics{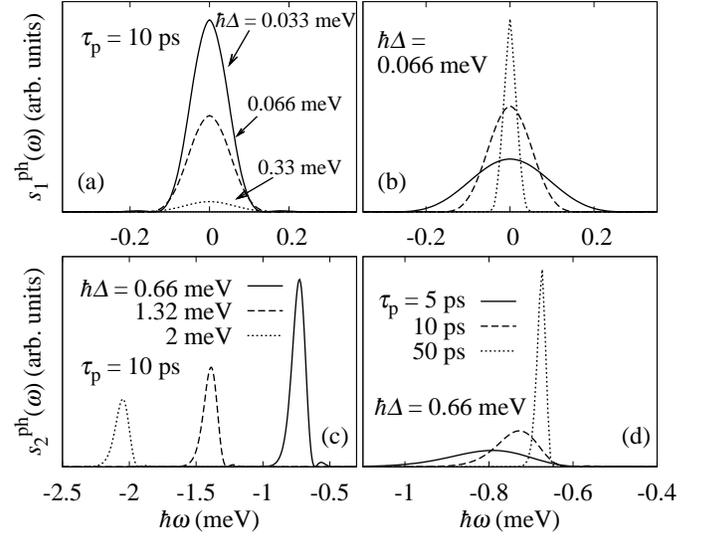}}}
\end{center} 
\caption{\label{fig:driving12} (a),(b) Spectral function $s_{1}^{\mathrm{ph}}(\omega)$.
(c),(d) Spectral function $s_{2}^{\mathrm{ph}}(\omega)$.
In (b), (d) the line styles denote pulse durations as defined in (d).}
\end{figure}

The symmetric function $s_{1}^{\mathrm{ph}}(\omega)$
[see Fig.~\ref{fig:driving12}(a,b)] is centered at $\omega=0$
and, for longer pulse durations, covers the low frequency part
(broadening $\approx 1/\tau_{\rm p}$).
The broadening of this function is independent of the detuning,
and, for a fixed $\tau_{\mathrm{p}}$, 
its area decreases for larger detunings.
This part of the spectral characteristics corresponds to pure dephasing effects
\cite{krummheuer02,forstner03,krugel05,alicki04,machnik04}.
The resulting error [Eq.~(\ref{error-ph})] will decrease
for longer pulse durations as well as for larger detunings.

The second part of the spectral characteristics 
$s_{2}^{\mathrm{ph}}(\omega)$  [see Fig.~\ref{fig:driving12}(c,d)]
is centered at $\omega \approx -\Delta $,
and its area grows with time.
Thus, the corresponding error contribution is proportional to the
spectral density of the phonon reservoir around the frequency
corresponding to the detuning of the laser frequency from the trion
transition, with some broadening due to time dependence. Moreover, the
error increases for longer operations. Therefore, 
this contribution may be interpreted as a real transition
and describes the error due to phonon-assisted trion generation.

In order to see this more directly, let us note that the function 
$s_{2}^{\mathrm{ph}}(\omega)$ is relatively strongly localized around
$\omega=-\Delta$, compared to
the range of variation of $R(\omega)$. Therefore, the corresponding
integral in Eq.~(\ref{error-ph}) may be approximated by its Markovian
limit,
\begin{displaymath}
\delta^{\mathrm{ph}}_{2}
=R_{\mathrm{ph}}(-\Delta)\int d\omega s_{2}^{\mathrm{ph}}(\omega).
\end{displaymath}
The area of the spectral function $s_{2}^{\mathrm{ph}}(\omega)$ appearing in
this formula may be calculated by noting that for any function $h(t)$,
one has
\begin{eqnarray}
\int d\omega
\lefteqn{ \left|
\int dt \; e^{i \omega t}h(t)\right|^{2}=} \nonumber \\
&&\int d\omega\int dt\int dt' e^{i \omega (t-t')}h^{*}(t)h(t')\nonumber\\
& & =2\pi \int dt \left| h(t)\right|^{2}. \label{s-area}
\end{eqnarray}
Thus, we find
\begin{displaymath}
\delta^{\mathrm{ph}}_{2}= 2 \pi \int dt \; R_{\mathrm{ph}}(-\Delta)\frac{1}{4}
\cos^{2}\frac{\vartheta}{2} \sin^{2} \left[2\phi(t)\right].
\end{displaymath}
The expression under the integral is exactly the Fermi's golden rule
formula for the probability that a transition from the state
$|a_{1}\rangle$ to $|a_{2}\rangle$ will take place during the time
$dt$ due to the (diagonal) carrier-phonon coupling given in
Eq.~(\ref{Hc-ph}). Since the state $|a_{2}\rangle$ becomes the trion
state after the laser pulse is switched off, this process indeed
represents a phonon-assisted transition to the trion state. 

\begin{figure}[tb] 
\begin{center} 
\unitlength 1mm
{\resizebox{85mm}{!}{\includegraphics{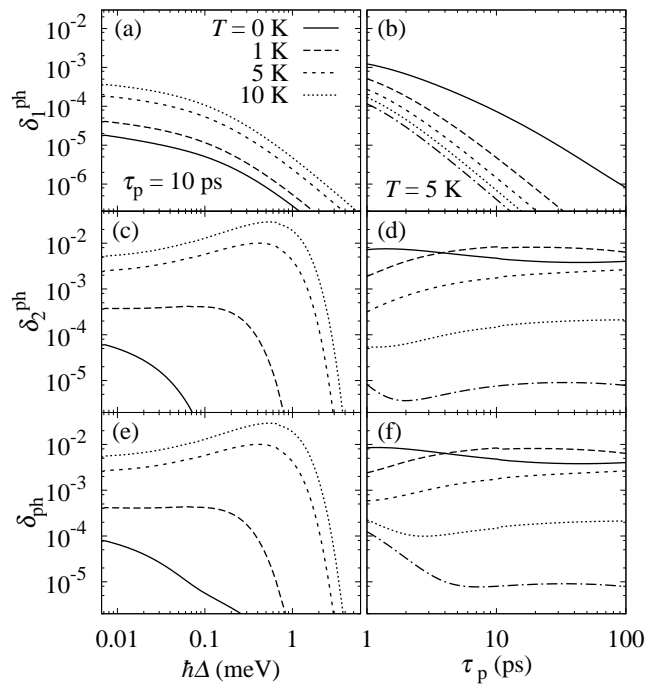}}}
\end{center} 
\caption{\label{fig:error-phonon} (a) The dependence 
of the error due to pure dephasing for growing detuning for
$\tau_{\rm p} = 10$ ps. (b) The pure dephasing error
as a function of pulse duration at $T = 5$ K and for five values
of detuning: $\hbar\Delta = 0.066$~meV (solid line),
$\hbar\Delta = 0.66$~meV (longer dashed line),
$\hbar\Delta = 1.32$~meV (shorter dashed line),
$\hbar\Delta = 2$~meV (dotted line),
and $\hbar\Delta = 2.64$~meV (dashed-dotted line).
(c), (d) As in (a), (b) but for the error due to phonon-assisted 
trion generation. (e), (f) As in (a), (b) but for the
total error resulting from the interaction of the carriers
with the phonon reservoir.
In (a), (c), and (e), the line styles denote temperatures as
defined in (a). In (b), (d), and (f), the lines styles denote
detunings as defined in (b).}
\end{figure}

The resulting phonon-induced errors, averaged over all initial states,
as functions of the detuning and pulse duration
are shown in Fig.~\ref{fig:error-phonon}.
The error due to pure dephasing $\delta_{1}^{\rm ph}$
[corresponding to $s_{1}^{\rm ph}(\omega)$]
favors longer pulse durations and larger detunings
and strongly depends on the temperature [Fig.~\ref{fig:error-phonon}(a,b)].
To perform the operation with an error smaller than $10^{-4}$,
which allows for coherent quantum operation on a qubit,
one needs a detuning larger than 0.06~meV (at $T = 5$~K).
It is possible to avoid the pure dephasing error even
for a fast evolution realized by pulse durations of several picoseconds
and a detuning of $1$~meV.

The contribution to the total error related to
the phonon-assisted trion generation $\delta_{2}^{\rm ph}$
[resulting from $s_{2}^{\rm ph}(\omega)$]
has a different behavior in comparison with the previous one
[Fig.~\ref{fig:error-phonon}(c,d)]
and depends even stronger on the temperature.
At low temperatures ($T<1$~K), it decreases with growing detuning,
but for higher temperatures ($T = 1$, $5$ and $10$~K),
it initially grows with detuning
and later decreases for considerably large detunings.
The maximum values correspond to the situation
when the spectral function $s_{2}^{\mathrm{ph}}(\omega)$
covers the maximum of the phonon density,
and the error vanishes if it lies beyond the phonon density cut-off.
The pulse duration dependence of this error differs from the one
for the pure dephasing error.
For relatively large detunings ($\hbar\Delta>1$~meV),
this error favors shorter pulse durations,
which is typical for real transitions.

In order to properly choose the conditions for the spin rotation
that lead to a high fidelity of the operation,
one needs to take into account these two sources of error
resulting from the carrier coupling to the phonon reservoir.
The sum of these two errors is plotted in Fig.~\ref{fig:error-phonon}(e,f).
To achieve values of the error below $10^{-4}$ for $\tau_{\rm p}=10$~ps,
detunings of several meV are needed.
For a small detuning ($\hbar\Delta=0.066$~meV), the error decreases with
growing pulse duration, but for larger detunings ($\hbar\Delta=0.66$~meV
and $\hbar\Delta=1.32$~meV), shorter pulse durations are more favorable.
Choosing the detunings larger than several meV can suppress
the influence of the phonon environment ($\hbar\Delta\gtrsim 4$~meV
at $T=5$~K).

\begin{figure}[tb] 
\begin{center} 
\unitlength 1mm
{\resizebox{85mm}{!}{\includegraphics{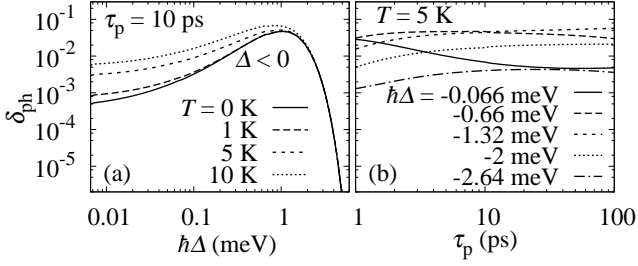}}}
\end{center} 
\caption{\label{fig:negative} The phonon-induced error for negative detunings
(a) at different temperatures for growing detuning and (b) for growing pulse duration.} 
\end{figure}

It is also possible to excite the system above the transition energies
by choosing a negative detuning.
In this case, the phonon-induced total error is larger in comparison
with the case for positive detuning (see Fig.~\ref{fig:negative}).
The corresponding spectral characteristics $s_{2}^{\mathrm{ph}}(\omega)$
is now centered at the positive frequency part of the phonon density,
which has larger values than the negative part, finite even at $T=0$~K
(see Fig.~\ref{fig:spectral-phonons}).
This corresponds to emission of a phonon,
which is strongly temperature dependent and possible at zero temperature.
Especially at low temperatures, the phonon-induced error 
is up to four orders of magnitude larger than the one for positive detunings.
To suppress the influence of the phonon reservoir, one needs larger detunings.
The dependence of this error on the pulse duration also differs.
Only for small detunings ($\hbar\Delta \lesssim 1$~meV), it is advantageous
to use longer pulses. For relatively large detunings ($\hbar\Delta > 1$~meV),
short pulses are favorable.

\subsection{Interaction with the photon field}

Since the considered procedure for spin rotation requires
a trion occupation during the evolution,
the radiative decay of the trion will result in an additional error.
We calculate the effect of the photon interaction
in the same manner as for the phonons
including perturbatively the carrier-photon interaction.

The relevant photon energies correspond to the semiconductor
band gap which is very large compared to the thermal energy. Therefore,
one can use the zero temperature approximation. 
The carrier-photon interaction contains the following
operators acting on the carrier subsystem:
$S_{02} = |0 \rl 2| e^{i\omega_{0}t}$, $S_{20} = S_{02}^{\dag}$,
$S_{12}=|1\rl 2|e^{i(\omega_{1}t-\gamma)}$, and $S_{21} =
S_{12}^{\dag}$.
All the resulting contributions, calculated according to the general
procedure in Sec.~\ref{sec:perturbation}, can be combined into a single 
spectral characteristics
\begin{equation*}
 S_{\rm rad}(\omega) = \sum_{i}\left| \langle \psi_{0}|
 Y^{\dag}_{\rm rad}(\omega)| \psi_{i} \rangle \right|^{2},
\end{equation*}
where
\begin{eqnarray*}
Y_{\rm rad}(\omega) & = & \frac{1}{\sqrt{2}}\int dt e^{-i\omega t}
U^{\dag}_{0}(t)\\
&&\times\left[  
\left(\cos\beta |B\rangle + \sin\beta |D\rangle
\right)e^{i\omega_{0}t}
\right.\\
&&+\left.\left(\sin\beta |B\rangle - \cos\beta |D\rangle \right)e^{i\omega_{1}t}
\right] \langle 2| U_{0}(t)
\end{eqnarray*}
and the sum is taken over all states
orthogonal to the initial state $| \psi_{0}\rangle$ [Eq.~(\ref{psi0})].

The spectral density of the photon reservoir reads \cite{scully97}
\begin{equation}\label{R-rad}
R_{\rm rad}(\omega) = 
\frac{|\vec d|^{2} \omega^{3}n_{\mathrm{r}}}{6 \pi^{2} \hbar \epsilon_{0} c^3},
\quad \omega>0,
\end{equation}
and the contribution to the error due to the photon interaction has the form
\begin{equation*}
\delta_{\rm rad} = \int d\omega R_{\rm rad}(\omega) S_{\rm rad}(\omega).
\end{equation*}

Using the definition (\ref{Y}) and the explicit form of the evolution
operator (\ref{Uan}), one finds 
\begin{equation*}
\langle \psi_{0}| Y^{\dag}_{\rm rad}(\omega)| \psi_{i} \rangle
= \frac{1}{\sqrt{2}}\int dt \; e^{i \omega t}s_{i}^{\mathrm{rad}}(t),
\end{equation*}
where 
\begin{eqnarray*}
\lefteqn{s_{1}^{\mathrm{rad}}(t)=}\\
&&- \frac{1}{4} \sin \vartheta \sin[2\phi(t)] 
(e^{-i \omega_{0} t}\cos \beta + e^{-i \omega_{1} t}\sin \beta ) \\ 
&& - e^{i[\varphi+\Lambda_{1}(t)]} \cos^{2} \frac{\vartheta}{2} \sin \phi(t)
(e^{-i \omega_{1} t}\cos \beta - e^{-i \omega_{0} t}\sin \beta ) 
\end{eqnarray*}
and
\begin{eqnarray*}
s_{2}^{\mathrm{rad}}(t) & = & 
- e^{-i[\Lambda_{2}(t) - \Lambda_{1}(t)]}
\cos \frac{\vartheta}{2} \sin^{2} \phi(t) \\
&&\times
(e^{-i \omega_{0} t}\cos \beta + e^{-i \omega_{1} t}\sin \beta ).
\end{eqnarray*}

The spectral function $S_{\rm rad}(\omega)$ is centered at
the laser frequency $\omega=\omega_{0,1} \approx 1$ eV,
and its width is of the order of 1 meV or less.
The photon spectral density $R_{\rm rad}(\omega)$
[Eq.~(\ref{R-rad})] is very broad and may be assumed constant
in the area of the overlap with $S_{\rm rad}(\omega)$.
Therefore, we use the Markovian approximation and obtain 
$\delta_{\rm rad} = R_{\rm rad}(\omega_{0}) \int d\omega S_{\rm rad}(\omega)$
with $R_{\rm rad}(\omega_{0})= \Gamma/(2\pi)$, where $\Gamma$ is the trion decay
rate $\approx 1$ ns$^{-1}$.
The frequency integral can be performed using Eq.~(\ref{s-area}).
Upon averaging over the initial states,
one obtains the resulting error due to the carrier-photon interaction:
\begin{widetext}
\begin{eqnarray}\label{error-rad}
\delta_{\rm rad}  & = & \Gamma \int dt \left(
\left\{\frac{1}{24}\sin^{2}[2\phi(t)]+\frac{1}{4}\sin^{4}\phi(t) \right\}
\left\{ 1 +\cos\left[(\omega_{1}-\omega_{0})t\right] \sin(2\beta) \right\}
\right. \\ \nonumber && \left. \phantom{aaaaaaa}
+\frac{1}{3}\sin^{2}\phi(t) \left\{ 1 -\cos \left[ (\omega_{1}-\omega_{0})t
 \right] \sin (2\beta)\right\}\right).
\end{eqnarray}
\end{widetext}

For large enough detunings ($\Omega \ll \Delta$), we can again obtain
an approximate equation for the resulting error:
\begin{eqnarray*}
\lefteqn{\delta_{\rm rad} \approx} \\ \nonumber
 && \Gamma \left( \frac{\sqrt{\pi} \Omega^{2}\tau_{\rm p}}{8 \Delta^{2}} \left\{
\frac{5}{3}-\exp{\left[-\frac{1}{4} \tau_{\rm p}^{2} (\omega_{1}-\omega_{0})^{2} \right]} \right\}
\right. \\ \nonumber && \left. \phantom{\frac{1}{1}}+
\frac{\sqrt{\pi} \Omega^{4}\tau_{\rm p}}{8\sqrt{2} \Delta^{4}}\left\{
5-\exp{\left[-\frac{1}{8} \tau_{\rm p}^{2} (\omega_{1}-\omega_{0})^{2} \right]} \right\}
\right).
\end{eqnarray*}

\begin{figure}[tb] 
\begin{center} 
\unitlength 1mm
{\resizebox{85mm}{!}{\includegraphics{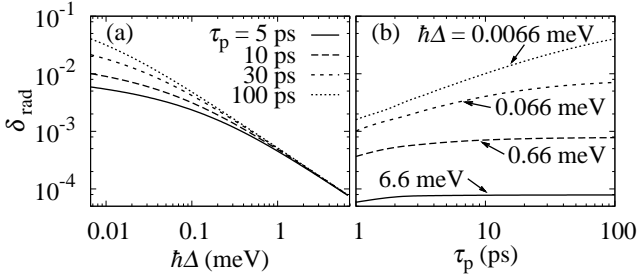}}}
\end{center} 
\caption{\label{fig:error-rad} The dependence 
of the radiative error for growing (a) detuning
and (b) pulse duration.}
\end{figure}

The resulting radiative error [Eq.~(\ref{error-rad})] as a function
of detuning and pulse duration is plotted in Fig.~\ref{fig:error-rad}(a,b).
This error decreases with growing detuning since the trion
occupation is reduced.
For small detunings, the trion occupation is relatively large
since the system is excited near the resonance,
and the resulting error is growing with pulse duration
due to the growing probability of the radiative decay of the trion.
In this regime, the error is linear in the pulse duration,
$\delta_{\rm rad} \approx \tau_{\rm p} \Gamma$.
The contribution to the error due to the finite trion lifetime
depends strongly on the pulse duration only for relatively small detunings.
For detunings of several meV, this error is constant in time.
To insure a small radiative error, one has to properly
choose a large detuning, while the pulse duration can be
arbitrarily short.


\section{Interplay of the different kinds of errors}
\label{sec:interplay}

In the previous sections, we studied the detuning
and pulse duration dependence of the contributions
to the total error of the spin-based quantum gate.
In this section, we calculate the resulting total error
and discuss the interplay of and possible optimization
against the constituent sources of the error.

\begin{figure}[tb]
\begin{center} 
\unitlength 1mm
{\resizebox{85mm}{!}{\includegraphics{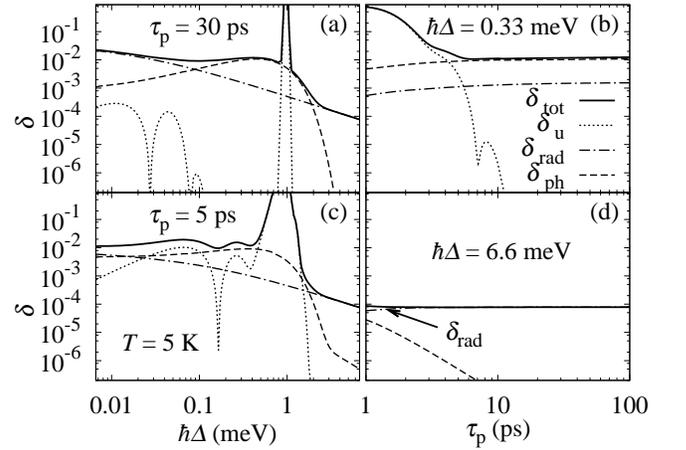}}}
\end{center} 
\caption{\label{fig:error-contr} Interplay of the error contributions
as a function of (a), (c) detuning and (b), (d) pulse duration at $T = 5$ K.
Different lines represent various error sources, as defined in (b).}
\end{figure}

We start the discussion with the dependence of the contributions
to the total error of the considered spin rotation on the detuning
for a fixed temperature ($T=5$~K) and two pulse durations
($\tau_{\rm p}=30$~ps and $\tau_{\rm p}=5$~ps) [see Fig.~\ref{fig:error-contr}(a,c)].
For small detunings, the trion occupation is large,
and the dominant source of the error is the radiative decay of the trion.
This contribution decreases with growing detuning,
and the phonon-induced error becomes the most important source of
dephasing. 
If the values of the detuning approach the Zeeman splitting,
one spin state is almost resonantly coupled to the trion state,
and the probability of the leakage to the trion state is high,
which leads to a large error due to the unitary corrections,
especially in the case of a short pulse ($\tau_{\rm p}=5$~ps).
In the detuning regime between $0.09$~meV and $2$~meV,
the error resulting from the phonon coupling becomes dominant
and is one order of magnitude larger than the radiative error.
For detunings of several meV
above the value of the cut-off of the phonon error,
the only contribution which inhibits the coherent spin rotation
is the trion radiative decay and has values between $10^{-3}$ and $10^{-4}$.
To achieve errors smaller than $10^{-4}$,
large detuning of several or tens of meV are needed.
In this detuning regime, the limitation results from the fact
that one cannot choose an arbitrarily large detuning,
in particular not with a frequency corresponding to
that of optical phonons, assumed to be well off-resonant in this paper.
Thus, the interplay of the different kinds of error
leads to a nontrivial detuning dependence of the total error,
which in general is dominated by the error due 
to the interaction with the phonon and photon reservoirs

Let us now discuss the dependence of the particular
error contributions on the pulse duration 
for a fixed temperature ($T=5$~K) [see Fig.~\ref{fig:error-contr}(b,d)].
For a small detuning ($\hbar\Delta=0.33$~meV)
and fast driving fields (short pulse durations of several picoseconds),
the adiabatic condition is not fulfilled, and the probability
of the leakage to the trion state becomes very high,
which results in large errors.
To minimize the influence of the errors due to the unitary corrections,
one has to choose pulse durations of at least a few picoseconds.
In this regime, the total error is limited by the phonon-induced contribution.
The second dominant source of the error is the radiative decay of the trion.
Moreover, these two contributions are almost constant in time.
For a large detuning ($\hbar\Delta=6.6$~meV), the evolution
is perfectly adiabatic, and the error due to unitary corrections does not appear.
The phonon-induced error is very small and vanishes for pulse durations
of a few picoseconds, since the detuning is far above the 
value of the cut-off of the phonon spectral density.
The only important contribution to the total error is due to
the finite trion lifetime which is constant in pulse duration.
One can see that it is advantageous to excite with larger detunings,
but the variation of the pulse duration is not of great importance
as soon as the adiabatic condition is met and the detuning
is not close to the Zeeman splitting.

\begin{figure}[tb]
\begin{center} 
\unitlength 1mm
{\resizebox{85mm}{!}{\includegraphics{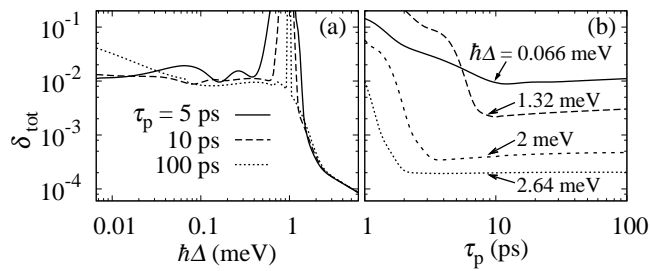}}}
\end{center} 
\caption{\label{fig:error-total} The total error
for growing (a) detuning and (b) pulse duration at $T = 5$ K.}
\end{figure}

In order to summarize the study of the various sources of decoherence,
we calculated the total error of the spin-based qubit rotation
for growing detuning [Fig.~\ref{fig:error-total}(a)]
and pulse duration [Fig.~\ref{fig:error-total}(b)].
For detunings smaller than $1.3$~meV, the error strongly depends
on the pulse duration and is relatively large.
Thus, to perform a rotation with a high fidelity,
detunings of several meV are needed.
Furthermore, for such detunings, the error is constant
for growing pulse duration, which opens the possibility
to perform the qubit rotation with pulse durations three orders of
magnitude smaller 
than the lifetime of the trion ($\Gamma^{-1} \approx 1$ ns).
For $\hbar\Delta\sim 5$~meV, the error may be as low as
$10^{-4}$ and is independent of the pulse duration.

\section{Conclusions}
\label{sec:conclusion}

We have studied a theoretical proposal of the optical control
of a single spin in a single doped quantum dot \cite{chen04}.
We have investigated the sources of error of a quantum operation
on a spin-based qubit and have given quantitative estimates
for the implementation of spin rotation through an angle
of $\pi/2$ about the $z$ axis in a self-assembled InAs/GaAs system.
The dephasing mechanisms resulting from the interaction of the carriers
with phonon and photon reservoirs as well as the imperfections
of the unitary evolution have been considered.

We have shown that the interplay of the constituent sources of the error
leads to a nontrivial dependence of the total error on the detuning,
which is in general dominated by the errors due to the coupling
of the carriers to the phonon and photon environments.
Furthermore, small detunings or detunings approaching the Zeeman splitting 
should be avoided since they lead to large trion occupations
which suppress an adiabatic and coherent control.
Taking into account all contributions to the total error,
we showed that 
errors as low as $10^{-4}$ can be achieved for 
large detunings ($\sim 5$~meV),
while the pulse durations can in principle be arbitrary
(but at least of a few picoseconds).

Finally, it should be noted that the calculations were performed using simple
Gaussian pulses with the pulse duration and intensity as the only
tunable parameters. Further reduction of the errors is very likely to
be possible with pulse optimization \cite{wenin06,hohenester04,axt05}.

\begin{acknowledgments}
We thank C. Emary and M. Richter for fruitful discussions.
A.G. acknowledges financial support from the DAAD.
This work was partly supported by Grant No. N20207132/1513 of the
Polish MNiSW. 
\end{acknowledgments}

\appendix

\section{Phonon couplings, spectral density}
\label{app:couplings}

In this Appendix, we derive the effective coupling element 
for the carrier-phonon interaction $F_{22}(\kk)$ as well as the resulting spectral density
of the phonon reservoir $R_{\rm ph}(\omega)$ for the studied QD system.

The general interaction Hamiltonian for confined states (restricted to
the ground states of the carriers) reads
\begin{eqnarray*}
H_{\mathrm{int}} & = & \sumk \sum_{\sigma} 
\left[D_{\mathrm{e}}\mathcal{F}_{\mathrm{e}}(\kk)
a_{\mathrm{e},\sigma}^{\dag} a_{\mathrm{e},\sigma}
-D_{\mathrm{h}}\mathcal{F}_{\mathrm{h}}(\kk)
 a_{\mathrm{h},\sigma}^{\dag} a_{\mathrm{h},\sigma}\right]\\
&& \times\sqrt{\frac{\hbar k}{2 \rho V c_{\rm l}}}
\left(b_{\kk}^{\phantom{\dag}} + b_{-\kk}^{\dag}\right),
\end{eqnarray*}
where $a_{\mathrm{e(h)},\sigma},a_{\mathrm{e(h)},\sigma}^{\dag}$ 
are the annihilation and creation operators, respectively,
for an electron or a hole in the
confined ground state with spin $\sigma$, and
the form factors are given by 
\begin{equation}
\mathcal{F}_{\mathrm{e(h)}}(\kk) = \int_{-\infty}^{+\infty} d^{3}\bm{r} \;
\Psi_{\mathrm{e(h)}}^{*}(\bm{r}) e^{i\bm{kr}} \Psi_{\mathrm{e(h)}}(\bm{r}) =
\mathcal{F}_{\mathrm{e(h)}}^{*}(-\kk).
\end{equation}
For Gaussian wave functions as in Eq.~(\ref{wavefunction}), one
explicitly finds by a simple integration
\begin{equation*}
\mathcal{F}_{\mathrm{e(h)}}(\kk) = 
\exp{\left(-\frac{1}{4} k_{\perp}^{2}l_{\mathrm{e(h)}}^{2}
-\frac{1}{4} k_{z}^{2}l_{z}^{2} \right)}.
\end{equation*}

For a single electron state 
$|0\rangle=a_{\mathrm{e},\uparrow}^{\dag}|\mathrm{vac}\rangle$ or 
$|1\rangle=a_{\mathrm{e},\downarrow}^{\dag}|\mathrm{vac}\rangle$ 
($|\mathrm{vac}\rangle$ is the empty dot state), one immediately finds
\begin{eqnarray*}
\langle 0|H_{\mathrm{int}}|0\rangle & = & 
\langle 1|H_{\mathrm{int}}|1\rangle \\
& = &
\sumk D_{\mathrm{e}}\sqrt{\frac{\hbar k}{2 \rho V c_{\rm l}}}
\mathcal{F}_{\mathrm{e}}(\kk)\left(\bk+\bmkd\right).
\end{eqnarray*}
For the trion state 
$|2\rangle=a_{\mathrm{e},\uparrow}^{\dag}
a_{\mathrm{e},\downarrow}^{\dag}
a_{\mathrm{h},\uparrow}^{\dag}|\mathrm{vac}\rangle$,
one has 
\begin{eqnarray*}
\lefteqn{\langle 2|H_{\mathrm{int}}|2\rangle=}\\
&&\sumk \sqrt{\frac{\hbar k}{2 \rho V c_{\rm l}}}
\left[ 2D_{\mathrm{e}}\mathcal{F}_{\mathrm{e}}(\kk)
-D_{\mathrm{h}}\mathcal{F}_{\mathrm{h}}(\kk) \right]
\left(\bk+\bmkd\right).
\end{eqnarray*}

In order to further simplify the calculations, we neglect the
difference between the localization sizes of the electron and hole
wave functions. Taking into account 
the small variation of the electron and hole confinement widths
$l_{\rm e}$ and $l_{\rm h}$ leads only to
inessential quantitative corrections \cite{grodecka06}.
This leads to the form factors 
$\mathcal{F}_{\mathrm{e}}(\kk)=\mathcal{F}_{\mathrm{h}}(\kk)
=\mathcal{F}(\kk)$ [Eq. (\ref{formfactor})] and to
Eq.~(\ref{hcph}) with 
$f_{00}=f_{11}= \sqrt{\frac{\hbar k}{2 \rho V c_{\rm l}}}D_{\mathrm{e}}
\mathcal{F}(\kk)$ and 
$f_{22}= \sqrt{\frac{\hbar k}{2 \rho V c_{\rm l}}}
(2D_{\mathrm{e}}-D_{\mathrm{h}})\mathcal{F}(\kk)$,
and thus to Eq.~(\ref{f22}).

With the isotropic acoustic phonon dispersion, we obtain the
spectral density of the phonon reservoir $R_{\rm ph}(\omega)$ [Eq.~(\ref{Rph})]:
\begin{eqnarray*}
R_{\rm ph}(\omega) & = & [n_{\rm B}(\omega)+1] \frac{V}{(2\pi)^{3}\hbar^{2}}	
\int_{0}^{2\pi} d\eta \int_{-\pi/2}^{\pi/2} d\zeta 
\cos\zeta  \\
&& \times  \int dk \; k^{2}\left| F_{22} (\kk)  \right|^{2} 
\left[ \delta(\omega-\wk) + \delta(\omega+\wk) \right],
\end{eqnarray*}
where the angles $\eta$ and $\zeta$ denote the orientation of the $\kk$ vector.
We can rewrite it in the form
\begin{equation*}
R_{\rm ph}(\omega) = R_{0} [n_{\rm B}(\omega)+1] \omega^{3} g(\omega),
\end{equation*}
where
\begin{equation*}
R_{0} = \frac{\hbar (D_{\rm e}-D_{\rm h})^{2}}{8 \pi^{2} \rho c_{\rm l}^{5}}
\end{equation*}
and the function $g(\omega)$ is defined as
\begin{eqnarray*}
\lefteqn{g(\omega)  =}\\ 
&& \int_{-\pi/2}^{\pi/2} d\zeta \cos\zeta 
\exp \left[ -\frac{l^{2}\omega^{2}}{2 c_{\rm l}^{2}} \left(
\cos^{2}\zeta + \frac{l_{z}^{2}}{l^{2}} \sin^{2}\zeta \right) \right].
\end{eqnarray*}

\section{Lindblad master equation for the trion recombination channel}
\label{app:lindblad}

In this Appendix, we derive the results for the trion recombination channel
in the Lindblad formalism and compare them with those calculated
by means of the perturbative method discussed in this paper.

From the interaction Hamiltonian $H_{\rm c-rad}$ [Eq.~(7)],
we derive the Lindblad equation \cite{breuer02} in the form
\begin{equation*}
\dot\rho = \Gamma \left( \sigma_{-} \rho \sigma_{+} - \frac{1}{2}
\sigma_{+}\sigma_{-} \rho -\frac{1}{2} \rho \sigma_{+}\sigma_{-} \right)
- i \left[ H_{\rm ad} , \rho \right],
\end{equation*}
where
\begin{eqnarray*}
\lefteqn{\sigma_{+} = } \\
&& \frac{1}{2}\left[ e^{-i\omega_{0}t} \left(
|2\rl B| + |2\rl D| \right) + e^{-i\omega_{1}t} \left(
|2\rl B| - |2\rl D| \right) \right]
\end{eqnarray*}
and $H_{\rm ad} = i \dot U_{\rm C}(t) U_{\rm C}^{\dag}(t)$
is the Hamiltonian generating the adiabatic evolution, 
with $U_{\rm C}$ given by Eq.~(\ref{Uan}). 
This equation is consistent with the perturbative approximation
in the sense that the latter is reproduced upon
transforming to the interaction picture and
performing an expansion in carrier-phonon coupling.

\begin{figure}[h]
\begin{center} 
\unitlength 1mm
{\resizebox{85mm}{!}{\includegraphics{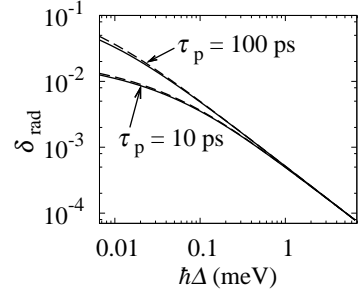}}}
\end{center} 
\caption{\label{fig:comparison} The radiative error calculated
by means of the Lindblad (solid lines) and perturbative (dashed lines)
methods for growing detuning.}
\end{figure}

The results from the Lindblad equation together with those
calculated with the perturative theory are plotted in Fig.~11 
for the initial state $|\psi_{0}\rangle = |B\rangle$
for the $\pi/2$ rotation about the $z$ axis.
For small detunings ($\hbar\Delta < 0.1$~meV),
where the error is relatively large,
the perturbative method yields slightly larger errors,
while in the case of larger detunings the results are the same.

\end{document}